\documentclass[pre,twocolumn,superscriptaddress,showpacs]{revtex4}

\usepackage{graphicx}

\begin{document}

\title{Diffusion on random site percolation clusters. Theory and NMR microscopy experiments with model objects}

\author{Andreas Klemm}
\affiliation{Sektion Kernresonanzspektroskopie, Universit\"{a}t Ulm, 89069 Ulm, Germany}
\author{Ralf Metzler}
\affiliation{Department of Physics, Massachusetts Institute of Technology,
77 Massachusetts Avenue, Room 12-109, Cambridge MA 02139, U.S.A.}
\author{Rainer Kimmich}
\affiliation{Sektion Kernresonanzspektroskopie, Universit\"{a}t Ulm, 89069 Ulm, Germany}

\begin{abstract}
Quasi two-dimensional  random site percolation model objects were
fabricated based on computer generated templates. Samples consisting of two
compartments, a reservoir of H$_2$O gel attached to a percolation model
object which was initially filled with D$_2$O, were examined with NMR
(nuclear magnetic resonance) microscopy for rendering proton spin density maps.
The propagating proton/deuteron inter-diffusion profiles were recorded and
evaluated with respect to anomalous diffusion parameters. The deviation of the
concentration profiles from those expected for unobstructed diffusion directly
reflects the anomaly of the propagator for diffusion on a percolation cluster.
The fractal dimension of the random walk, $d_w$, evaluated from the diffusion
measurements on the one hand and the fractal dimension, $d_f$, deduced from the
spin density map of the percolation object on the other permits one to
experimentally compare dynamical and static exponents. Approximate calculations
of the propagator are given on the basis of the fractional diffusion equation. Furthermore, the
ordinary diffusion equation was solved numerically for the corresponding initial and boundary
conditions for comparison. The anomalous diffusion constant was evaluated and is compared to the Brownian
case. Some ad hoc correction of the propagator is shown to pay tribute to the
finiteness of the system. In this way, anomalous solutions of the fractional
diffusion equation could experimentally be verified for the first time.
\end{abstract}

\pacs{05.40.-a, 82.56.Lz, 47.53.+n, 64.60.Ht}

\maketitle

\section{Introduction}

Randomly disordered media are present in many fields of nature and science. The
dynamical properties ruled by the geometrical structure are of special interest
in fields of physical and engineering processes like filtering and exploration
\cite{sahimi95,dullien92,adler98}. Percolation theory has proven to be a
powerful tool to model porous systems \cite{stauffer92,sahimi93,bunde96}.

 The objective of this study is to examine diffusion on random site
percolation clusters experimentally and analytically. There are several
numerical simulation studies in the literature suggesting an anomalous
displacement behavior related to the fractal nature of the clusters
\cite{orbach86,levitz97,havlin87}. However, there is little experimental
evidence for the reality and practical detectability of anomalous diffusion so
far \cite{klammler92,kimmich93,kimmich95}.

The objective of the present work is to exploit a new experimental strategy.
This is: (a) to generate numerically a percolation model cluster, (b) to
determine the characteristic parameters numerically, (c) to fabricate model
objects using the percolation clusters as templates, (d) to record nuclear
magnetic resonance (NMR) spin density maps from the (water-filled) pore space,
(e) to evaluate the characteristic cluster parameters on this basis again, (f)
to study interdiffusion of heavy and light water in the pore space, and (g) to
compare the experimental interdiffusion profiles with solutions of the
fractional diffusion equation \cite{metzler00} for the first time. In a sense,
we are thus continuing our previous work in which we had already explored
static and dynamic properties in various three-dimensional and quasi
two-dimensional percolation model objects
\cite{muller95,muller96,klemm97,klemm00}.

Random site percolation structures are defined in the  two-dimensional case by
sites on a square lattice. They are occupied with a probability $p$ that is
usually chosen in the vicinity of the (two-dimensional) percolation threshold,
$p_c= 0.592746$. Neighboring occupied sites are connected by pores with a
cross-section corresponding to the lattice constant or integer multiples of it.
The total subset of connected lattice sites form a so-called cluster. For
$p\geq p_c$, sample-spanning clusters occur that can be examined with respect
to transport properties. The pore space structure generated by the random site
percolation model can be characterized by four parameters, that is, the lattice
constant, $a$, the fractal dimension, $d_f$, the correlation length, $\xi$, and
the percolation probability, $P_\infty$ \cite{stauffer92,bunde96,herrmann91}.
The latter quantity is defined as the probability that a site belongs to the
``infinite'' cluster traversing the whole sample \cite{kapitulnik83}. The
correlation length, which is of particular interest here, is defined as the
mean distance between two sites of a finite cluster (or the mean hole diameter
in an infinite cluster). In the real percolation model objects we are
considering here, the minimum lattice constant (or pore diameter) is given by
the mechanical resolution of the fabrication process (see below).

Random site percolation clusters are known  to display fractal properties on a
length scale below the correlation length. That is, the volume-averaged
porosity scales with the probe volume radius, $r_p$, as
\begin{equation}
\rho \propto
\left\{ \begin{array}{l}
 r_p^{d_f-d_E} \hspace*{5mm} \mbox{for} \hspace*{5mm} a \ll r_p \ll \xi \\
P_\infty \propto r_p^0 \hspace*{5mm} \mbox{for} \hspace*{5mm} r_p \gg \xi .
\end{array} \right.
\label{rho}
\end{equation}
The Euclidean dimension is  denoted by $d_E$ ($=2$ in the present case). The
fractal dimension for $d_E=2$ was theoretically derived as $d_f=91/48 \approx
1.896$ \cite{stauffer92}.

The purely structural relationship Eq.~(\ref{rho}) is opposed by the dynamic
property for the mean squared displacement of a random walker on the cluster,
\begin{equation}
\left\langle r_d^2 \right\rangle \propto
\left\{ \begin{array}{l}
t^{2/d_w} \hspace*{5mm} \mbox{for} \hspace*{5mm}t_a \ll t \ll t_\xi \\
D_{eff} \, t  \hspace*{5mm} \mbox{for} \hspace*{5mm} t \gg t_\xi ,
\end{array} \right.
\label{msd}
\end{equation}
where $t_\xi \propto \xi^{d_w}$ is the  time the random walker needs to explore
the correlation length $\xi$, and $d_w$ is the fractal dimension of the random
walk. The lower time limit of the anomalous diffusion regime is given by the
time needed for the displacement of length $a$, $t_a$. The diffusion
coefficient becoming effective in the long-time limit, $t \gg t_\xi$, is
denoted by $D_{eff}$. According to the Alexander/Orbach conjecture
\cite{alexander82}, the quantity $d_w$ is assumed to be related to the fractal
dimension as
\begin{equation}
d_w = \frac{3}{2} d_f \hspace*{5mm} \mbox{ for } \hspace*{5mm} d_E \geq 2.
\label{aoc}
\end{equation}
That is, the structural parameter $d_f$ characteristic for the volume-averaged
porosity is linked to the dynamic parameter $d_w$ specifying anomalous
diffusion. For $d_E=2$, the diffusion exponent becomes $d_w \approx 2.87$. In
this study experimental evaluations for both quantities have been carried out,
so that a comparison becomes possible. Note, however, that Eq.~(\ref{aoc}) is
not considered to be an exact relation
\cite{stauffer92,bunde96,hong84,zabolitzky84}.

A theoretical problem of intriguing impact is the complete propagator
description of anomalous diffusion rather than restricting oneself to the
second moment of the propagator according to Eq.~(\ref{msd}). In the second
part of this article, the analytical treatment based on the fractional
diffusion equation \cite{metzler00,schneider} is outlined and compared with the
inter-diffusion profile data acquired in our experiments.

\section{Techniques and instruments}

\subsection{Methods for measuring diffusion}

In the model objects to  be studied here, the minimum pore diameter is $\Delta
r= 400 \, \mu$m. The displacement length scale needed to probe anomalous
diffusion, i.e. displacements obstructed by the matrix, is $r_d \gg \Delta r$.
The ordinary pulsed gradient spin echo technique (see
Refs. \cite{kimmich97,callaghan91}, for instance) is therefore not
suitable for the detection of anomalies in liquids in the present situation.

The much larger displacement  rate in gaseous phases would permit such studies
in principle. In Ref. \cite{muller96} we have studied diffusion of
methane gas in a percolation model object. Although there was some indication
of an anomalous behavior, the experiment turned out to be difficult because of
the poor detection sensitivity. In this respect, diffusion studies using
laser-polarized or thermally polarized $^{129}$Xe are more promising
\cite{pasquier96,mair99}. Also, the use of inert fluorinated gases possibly at
somewhat elevated pressures may be more favorable \cite{lizak91,kuethe98}. In
any case, there is a diffusion mechanism (``Knudsen diffusion'') relevant in
gases which is different by nature from diffusion in liquids \cite{levitz97}.

We therefore preferred to  employ an isotope interdiffusion method. The samples
consisted of two compartments initially filled with H$_2$O (in gel form) and
D$_2$O. At the beginning of the experiment the compartments were pressed on
each other in close contact, so that interdiffusion was initiated. The time
evolution of the proton spin density maps in the D$_2$O compartment was then
studied as a function of time. The technique was ordinary one- or
two-dimensional NMR imaging of concentration profiles, an application for
diffusometry purposes already described in
Refs. \cite{heink78,kimmich88,klemm99}.

The proton density profiles  were recorded either in the form of
one-dimensional spin density maps, or were evaluated from the two-dimensional
spin density maps by projection on the main diffusion direction. The latter
variant has the advantage that signal noise from matrix areas can be screened
off before evaluating the profile data.

The propagation of the proton  density profile at half height as a function of
time permits one to determine the time dependence of the mean squared
displacements. Alternatively, the profiles themselves at a given time can be
examined with respect to the character of the diffusion process. In the latter
case, the full propagator characteristics and not just its second moment are
mattering.

\subsection{NMR tomograph and acquisition parameters}

The one- or two-dimensional  proton density maps of the water filled pore space
of percolation model objects were recorded with the aid of a NMR tomograph
consisting of a 4.7 T Bruker magnet with 40 cm horizontal room temperature bore
and a home made radio frequency console. Typical radio frequency and field
gradient pulse schemes for spin-echo NMR imaging can be found in
Ref. \cite{kimmich97}, for instance.  The spatial resolution of the
images was better than 300 $\mu$m. The acquisition of a two-dimensional spin
density map typically took 20' to 60', so that a reasonable time resolution was
given.

Isotopic dilution by deuterons prolongs  the local transverse and longitudinal
relaxation times due to the reduced number of dipolar interaction partners (see
Ref. \cite{fung77}, for instance). For the evaluation of spin density
maps, the spin echo signals therefore have to be corrected if the repetition
time is not much longer than the longest proton spin-lattice relaxation time,
$T_1$,  or if the echo time is not much shorter than the transverse relaxation
time, $T_2$.

Typical echo times, $T_E$, were between $20$ and  $30$ ms. This is to be
compared with transverse relaxation times of several seconds in water at room
temperature. Signal attenuation on this basis is therefore totally negligible.

The situation is less clear with  the effect of spin-lattice relaxation. The
repetition time, $T_R$, typically was 2~s, so that the spin density profiles
could be distorted at the low-concentration side by saturation effects. In some
of the experiments, we have therefore varied the repetition time between 0.25
and 12.2~s in order to evaluate the local spin-lattice relaxation times. The
local signal intensities were then corrected correspondingly to provide the
true spin density profiles. No significant spin-lattice relaxation effect could
be diagnosed (see the data discussed below).

\subsection{Computer generated percolation clusters}

In the insets of Fig.~\ref{fig7} and Fig.~\ref{fig9}, typical  two-dimensional
random site percolation clusters generated on a square lattice are shown. The
occupation probability, $p$, is slightly above the percolation threshold value
$p_c = 0.5927$ for the Euclidean dimension $d_E=2$ \cite{jan99}. The
volume-averaged porosity was evaluated using the so-called sandbox method
\cite{muller95,muller96}: $N_p$ probe circles of varying radius $r_p$ are first
placed randomly at positions \mbox{\boldmath$r_k$} within the cluster in such a
way that the center of the probe volume (which actually is an area in the
two-dimensional case) is in the pore space. Then the average values of the
observables are formed for the $N_V$ voxels at positions \mbox{\boldmath$r_j$}
inside the probe volume. Finally, the arithmetic mean of the data set for the
$N_p$ probe volumes with a given radius $r_p$ is taken. In other words, the
volume-averaged porosity is defined as
\begin{equation}
\rho(r_p)= \frac{1}{N_p}\sum_{k=1}^{N_p}\frac{1}{N_V}\sum_{j=1}^{N_V}\rho(\mbox{\boldmath$r_j$}),
\label{porosity}
\end{equation}
where $r\geq |\mbox{\boldmath$r_k$}- \mbox{\boldmath$r_j $}|$. This  quantity
can also be evaluated from black-and-white converted, experimental spin density
maps as described in Ref. \cite{muller95}.

\subsection{Model objects and measuring conditions}

The percolation model objects were  fabricated using a circuit board plotter
(for details see Refs. \cite{muller95,muller96,klemm97}) based on the
computer-generated templates. The mechanical fabrication resolution was $\Delta
r= 400 \, \mu$m (see photograph in Fig.~\ref{fig2}). The adjusted milling depth
ranged from 1 to 2 mm in the different objects produced.

The objects were filled with heavy  water and brought into contact with
reservoirs of H$_2$O gel (Kelcogel, 1.5~\% by weight) at time $t=0$, when the
interdiffusion process was to begin. The reservoirs are schematically shown in
the insets of  Fig.~\ref{fig7} and Fig.~\ref{fig9}. The proton distribution in
the objects was then measured as a function of time in the form of spin density
maps as described above.

The gel form of the undeuterated moiety  of the sample was needed for
stabilization and in order to prevent flow. On the other hand, gel
stabilization inside the percolation matrix is unfavorable because of the
tendency of forming voids upon gelation. It turned out that in this case the
stabilizing effect of the solid matrix is sufficient.

The influence of deuteration and  gelation on the bulk water self-diffusion
coefficient was checked in an ordinary pulsed-gradient spin echo experiment
\cite{kimmich97}. At 20~$^\circ$C and a gel content of 1.2~\%, the
self-diffusion coefficients of H$_2$O and D$_2$O were found to be $1.8\times
10^{-9}$~m$^2$/s and $1.4\times 10^{-9}$~m$^2$/s, respectively. That is, the
water self-diffusion coefficient is slightly reduced both by deuteration and
gelation.

An isotope effect is known to show  up already without gelation: Mills
\cite{mills73} reports self diffusion coefficients in pure H$_2$O and D$_2$O at
25~$^\circ$C $D_{H_2O}=2.3\times 10^{-9}$~m$^2$/s and $D_{D_2O}=1.87\times
10^{-9}$~m$^2$/s, respectively. That is, smaller diffusion coefficients are
expected with increasing dilution of $^{1}$H as it occurs at the diffusion
front.

Furthermore, there may be a difference  in the chemical potential in pure
H$_2$O and pure D$_2$O, so that a slight deviation from the self-diffusion
situation may play a role. On the other hand, no significant influence on the
shape and time evolution of the inter-diffusion profiles was found in the
bulk-to-bulk experiments described below. The conclusion therefore is all these
effects are of minor importance or largely compensate each other so that
(partial) deuteration and gelation does not perceptibly affect the percolation
cluster characteristics of the propagator to be probed.

The quasi two-dimensional model objects  were kept in a horizontal position
during the whole measuring process in order to avoid any convective
displacements. In order to improve the detection sensitivity, several identical
slices were stacked on each other (see Fig.~\ref{fig2}). All experiments were
carried out at room temperature, ($21 \pm 1$)~$^\circ$C.

\section{Results}

\subsection{Volume-averaged porosity}

The volume-averaged porosity was evaluated as a function of the probe volume
radius on the basis of Eq.~(\ref{porosity}) for the percolation clusters shown
in the insets  of Fig.~\ref{fig7} and Fig.~\ref{fig9}. Corresponding
evaluations from experimental spin density maps lead to equivalent decays as
demonstrated in Ref. \cite{klemm00}. The fractal dimension according to
Eq.~(\ref{rho}) was found to be $d_f=1.87$ in both cases. This value will be
compared  with the experimental value for $d_w$ according to Eq.~(\ref{aoc}).

\subsection{Isotope concentration profiles for interdiffusion between bulk water compartments}
\label{normal}

In order to test our measuring and evaluation  technique, we have recorded
isotopic inter-diffusion profiles in a sample consisting of two equal
compartments initially filled with bulk H$_2$O and D$_2$O gels, respectively.
In this case, the initial distribution of the proton density is of the type
\begin{equation}
 C(x,t=0) = \left\{
\begin{array}{c}
 C_0 \hspace*{5mm} \mbox{for} \hspace*{5mm} x\leq 0 \\
0 \hspace*{5mm} \mbox{for} \hspace*{5mm} x>0
\end{array}
\right. .
\end{equation}
Provided that the diffusion process is normal, that is,  in the absence of
obstructions by a matrix, the proton spin density profiles at later times are
given by \cite{crank75}
\begin{equation}
C(x,t) = \frac{ 1}{2} C_0 \mbox{ erfc} \left\{ \frac{ x}{2\sqrt{Dt}}\right\},
\label{profile}
\end{equation}
where $D$ is the diffusion coefficient and ${\rm erfc}(z)$ is the complementary
error function.  In principle, this solution of the ordinary diffusion equation
applies to ``infinite'' systems, that is, to root mean squared displacements
much less than the extension of the sample.

Figure~\ref{fig3} shows experimental proton spin  density maps measured in a
two-compartment sample (for an illustration see inset of Fig.~\ref{fig4}) as a
function of time $t$ after bringing the compartments into contact with each
other. The right and left compartments were initially filled with bulk H$_2$O
and D$_2$O gels, respectively. The mean proton spin density profiles, i.e.
projections of the two-dimensional spin density maps on the main diffusion
coordinate axis, $x$, are also shown in diagram form (white lines).

In Fig.~\ref{fig4}, these experimental concentration profiles are compared with
those predicted by Eq.~(\ref{profile}). The fits of Eq.~(\ref{profile}) to the
experimental data reproduces the room temperature value of the diffusion
coefficient measured in bulk water with the pulsed gradient spin echo technique
\cite{mills73,holz91}, $D=2\times 10^{-9}$~m$^2$/s, very well with the
exception of the shortest and longest diffusion intervals.

At the shortest diffusion interval, imperfections of the initial isotope
distribution and the limited time resolution of the (two-dimensional imaging
process) are expected to matter. The concentration profiles at the longest
diffusion times are already affected by the finite extension of the sample
which conflicts with the assumption in Eq.~(\ref{profile}) of infinite
compartments.

Taking together the potential sources of systematic  experimental errors
mentioned before, one can state that the agreement between the theoretical and
experimental concentration profiles in bulk samples is very reasonable so that
reliable evaluations of anomalous diffusion features in the percolation model
objects can be expected. This is corroborated by the determination of the mean
square displacement of the diffusion front in a second experiment. The set-up
is schematically shown in the inset of Fig.~\ref{fig5}. In this case, the
half-height positions of the concentration profiles (projections of the
two-dimensional spin density maps on the main diffusion direction) were
evaluated, squared and plotted versus time.

The middle section of the experimental curve  shown in Fig.~\ref{fig5} can
nicely be described by a power law $x_{1/2}^2  \propto t^{1.05}$ which is very
close to the linear mean squared displacement law for normal diffusion. As in
the experiment discussed before, the deviations at short times reflect the
initial situation that can only imperfectly be described by a step function.
The plateau reached at long times is due to the finite extension of the sample.

\subsection{Isotope concentration profiles for interdiffusion between bulk water and water filled percolation cluster compartments}
\label{anomalous}

Equation (\ref{profile}) is based on a Gaussian propagator as expected for
ordinary diffusion. That is, it does not account for diffusion in a percolation
cluster where the second moment of the propagator obeys an anomalous-diffusion
law as given in Eq.~(\ref{msd}).

Figure~\ref{fig6} shows two-dimensional proton spin  density maps acquired in
an experimental set-up schematically shown in the inset of Fig.~\ref{fig7}.
Inter-diffusion between a H$_2$O gel filled reservoir and a stack of quasi
two-dimensional percolation model objects was examined. Superimposed to the
spin density maps the mean spin density profiles (i. e.  projections on the $x$
axis) are shown in Fig.~\ref{fig6} as white lines.

The half-height positions of these mean concentration profiles, $x_{1/2}$, were
evaluated, squared and plotted versus time as shown in Fig.~\ref{fig7}. In the
limit of long diffusion times when the imperfections of the initial isotope
distribution and the finite time resolution of 26' do not matter anymore, the
data can be described by a power law again. The imperfection of the initial
isotope distribution is matched by the revised origin of time $t$.  Fitting the
exponent of Eq.~(\ref{msd}) for $t \ll t_\xi$ to the data leads to $d_w=2.89$
which favorably compares to the Alexander/Orbach conjecture, Eq.~(\ref{aoc}),
with angleichen the fractal dimension $d_f$ determined from the very same
percolation cluster according to Eq.~(\ref{rho}) for $a \ll r_p \ll \xi$.

The isotopic interdiffusion profiles can also be obtained directly with the aid
of one-dimensional imaging along the main diffusion direction. Fig.~\ref{fig8}
renders typical profiles recorded in this way from an experimental set-up
schematically shown in the inset of Fig.~\ref{fig9}. The data are corrected for
saturation effects due to incomplete spin-lattice relaxation after the
repetition interval. From these profiles, the mean squared proton displacement
(in the percolation cluster moiety) can be evaluated as a function of the
diffusion time. Figure~\ref{fig9} shows corresponding data. Details are
described in the legend. With a fitted exponent parameter of $d_w=2.86$ the
anomalous diffusion character in the percolation cluster is again corroborated
in good agreement with the Alexander/Orbach conjecture Eq.~(\ref{aoc}).

\section{Analytical propagator treatment based on the fractional diffusion equation}
\label{prop}

The propagator (or Green's function) for a Brownian random walk process in
one dimension is, of necessity, given in terms of the Gaussian
\begin{equation}
\label{gauss} C(x,\tau)=\frac{1}{\sqrt{4\pi \tau}}\exp\left(-\frac{x^2}{4\tau
}\right),
\end{equation}
due to the central limit theorem \cite{levy}. Expression~(\ref{gauss}) includes
the $\delta$-initial condition $C(x,0)=\delta(x)$ and fulfills natural boundary
conditions $\lim_{|x|\to \infty}C(x,t)=0$. We use the rescaled time $\tau\equiv
Dt$.  The propagator (\ref{gauss}) satisfies the diffusion equation \cite{levy}
\begin{equation}
\label{de} \frac{\partial}{\partial \tau}C=\frac{\partial^2}{\partial
x^2}C(x,\tau).
\end{equation}

Representing the experimental set-up shown in the inset of Fig.~\ref{fig5} by
the initial condition $C(x,0)=0$ for $x>0$ and through the boundary condition
$C(0,\tau)=C_0$, i.e., we assume that, due to the comparatively large
reservoir, a constant concentration is kept at the boundary $x=0$. The solution
of Eq.~(\ref{de}) for these initial and boundary conditions is given through
\begin{equation}
\label{csol} C(x,\tau)=C_0{\rm erfc}\left(\frac{x}{2\sqrt{\tau}}\right), \,\,
x>0.
\end{equation}
The solution given in Eq.~(\ref{csol}) reaches the plateau $C(x,\tau)=C_0$ for
$x^2\ll\tau$. Consequently, the expression in Eq.~(\ref{csol}) is not
normalized. Its mean,
\begin{equation}
\label{mean} \langle
1(\tau)\rangle\equiv\int_0^{\infty}C(x,\tau)dx=2C_0\sqrt{\tau/\pi},
\end{equation}
grows with the square root in time, and is proportional to the concentration
$C_0$. The second moment becomes
\begin{equation}
\langle x^2(\tau)\rangle=\frac{8}{3}C_0\tau^{3/2}/\sqrt{\pi}.
\end{equation}
In normalized form, that is, the concentration profiles are divided by the mean
given in Eq.~(\ref{mean}), the mean squared displacement corresponds to
\begin{equation}
\label{nmsd}
\langle x^2(\tau)\rangle_n=\frac{\langle x^2(\tau)\rangle}{\langle 1(\tau)
\rangle}=\frac{4}{3}\tau.
\end{equation}

It should be noted that the concentration profile, $C(x,\tau)$, for the
different realization with an infinite reservoir, Eq.~(\ref{profile}), differs
from the case with constant boundary concentration, Eq.~(\ref{csol}), only by a
factor $\frac{1}{2}$. Systems with constant boundary concentration and infinite
reservoir thus behave congruently.

Let us now address how we can derive the analogous quantities for the anomalous
diffusion data described in Section~\ref{anomalous}. As mentioned above, the
average diffusion in the percolation cluster close to the percolation threshold
is anomalous in the sense that the mean squared displacement follows $\langle
x^2(\tau)\rangle\propto \tau^{\alpha}$ where $\alpha=2/d_w$ [see
Eq.~(\ref{msd})], and $d_w=2d_f/d_s$ \cite{havlin}. Here, the rescaled time
$\tau=(D_{\alpha})^ {1/\alpha}t$ includes the anomalous diffusion constant
$D_{\alpha}$ of dimension $[D_{\alpha}]={\rm m}^2/{\rm s}^{\alpha}$
\cite{metzler00,rem}. Thus, the diffusion dynamics is controlled by two
parameters, the fractal dimension $d_f$ and the spectral dimension $d_s$.
Usually, the inequality $d_w>d_f$ is fulfilled so that the anomalous diffusion
process is {\em subdiffusive}, i.e., $0< \alpha<1$. In the experiment, the
three-dimensional probability density function is projected onto a
one-dimensional probability density function, and therefore the geometry in the
pseudo-one-dimensional process becomes averaged.

In the original geometry, the spreading of the random walker in space is slowed
down in comparison to the free diffusion, due to the presence of bottlenecks
and dead ends on all length scales, leading to the subdiffusive nature of the
mean squared displacement. In the projection, the trapping in a small pore of
the percolation cluster corresponds to a small wiggling around some given
coordinate. Effectively, the walker in the pseudo-one-dimensional measurement
experiences a continued multiple trapping process, i.e., it is immobilised for
some ``time'' span $\tau$ governed by the so-called waiting ``time''
distribution $\psi(\tau)$. According to this $\psi$, the walker is released and
moves freely until it is trapped again. This is a special case of a continuous
time random walk process, the subdiffusion being reflected in the inverse
power-law form $\psi(\tau)\sim A_{\alpha}\tau^{-1-\alpha}$ \cite{klafter}.
According to this $\psi$, the walker is released and moves freely until it is
trapped again, and so forth. This stop-and-go process can be mapped onto the
fractional diffusion equation \cite{metzler00,schneider}
\begin{equation}
\label{fde}
 \frac{\partial}{\partial \tau}C=\,
_0{\mathcal{D}} _{\tau}^{1-\alpha}\frac{\partial^2}{\partial x^2}C(x,\tau)
\end{equation}
that includes the Riemann-Liouville fractional operator
$_0{\mathcal{D}}_{\tau}^{1-\alpha} \equiv\partial/\partial \tau \big(\,_0
{\mathcal{D}}_{\tau}^{-\alpha}\big)$ with \cite{oldham}
\begin{equation}
\label{rl} _0{\mathcal{D}}_{\tau}^{-\alpha}C(x,\tau)
\equiv\frac{1}{\Gamma(\alpha)} \int_0^{\tau} d\tau'
\frac{C(x,\tau')}{(\tau-\tau')^{1-\alpha}}.
\end{equation}
The fractional diffusion equation is equivalent to a generalized master
equation, and can be derived from a multiple trapping version of the
fundamental Chapman-Kolmogorov equation \cite{gcke}.

The Riemann-Liouville operator has the important property that its Laplace
transform is $\int_0^{\infty}e^{-u\tau}\,
_0{\mathcal{D}}_{\tau}^{-\alpha}f(\tau)d\tau =u^{-\alpha}f(u)$. By virtue of
this property, it can be shown that in Laplace space, the solution of the
fractional diffusion equation, let us call it $C_{\alpha}(x,\tau)$, is
connected to the solution of the Brownian diffusion equation (\ref{de}),
$C_B(x,\tau)$, through the scaling relation $C_{\alpha}(x,u)=u^{\alpha-1}
C_B(x,u^{\alpha})$ \cite{metzler00}. This, in turn, can be used to find a
convenient mapping between the Brownian solution, and {\em any} quantity
obtained from it through linear operations, in terms of the generalized Laplace
transformation \cite{basil}
\begin{equation}
\label{genlap} C_{\alpha}(x,\tau)=\int_0^{\infty}E_{\alpha}(s,\tau)C_B(x,s)ds.
\end{equation}
In Eq. (\ref{genlap}), the function
$E_{\alpha}(s,\tau)$ is a one-sided L{\'e}vy distribution that can be
represented in terms of Fox's $H$-function \cite{metzler00}. For our
purposes, we notice that $E_{\alpha}$ has the convenient representation
\begin{eqnarray}
\nonumber
E_{2/3}(s,\tau)&=&\frac{1}{\tau^{2/3}\Gamma(1/3)}\, _1F_1\left(\frac{5}{6};
\frac{2}{3};-\frac{4s^3}{27\tau^2}\right)\\
&&-\frac{1}{\tau^{4/3}\Gamma(-1/3)}
\, _1F_1\left(\frac{7}{6};\frac{4}{3};-\frac{4s^3}{27\tau^2}\right)
\label{erep}
\end{eqnarray}
for $\alpha=2/3$. This value for $\alpha$ is sufficiently close to the
percolation cluster value $0.7$, and we choose $2/3$ as for this rational
number the representation in Eq.~(\ref{erep}) considerably reduces the
computation time in evaluating the integrals of Eq.~(\ref{genlap}). In
Eq.~(\ref{erep}), $_1F_1$ represents the confluent hypergeometric function.

For the normalized mean squared displacement, Eq.~(\ref{nmsd}), the
transformation given in Eq.~(\ref{genlap}) yields the exact anomalous behaviour
\begin{equation}
\langle x^2(\tau)\rangle_n=\frac{2}{\Gamma(2/3)}\tau^{2/3}
\end{equation}
with $\tau\equiv (D_{2/3})^{3/2}t$ \cite{metzler00}. By virtue of this
expression, we can approximate the anomalous diffusion constant $D_{2/3}$ as
\begin{equation}
D_{2/3}\approx 1.3\times 10^{-8}\frac{{\rm m}^2}{{\rm s}^{2/3}}
\end{equation}
from the experimental data plotted in Fig.~\ref{fig9} where the observed time
interval approximately fulfills the boundary condition $C(0,\tau)=C_0$. To our
knowledge, this is the first time the anomalous diffusion constant appearing in
the above formalism has been determined experimentally. In the following, we
use $D_{2/3}$ to construct the anomalous concentration profile from
Eqs. (\ref{csol}) and (\ref{genlap}).

In Fig.~\ref{fig10}, we show the anomalous concentration profile $C_{\alpha}
(x,\tau)$ for the longest diffusion time interval ($\approx 180$~h) in
Fig.~\ref{fig8}, and compare it to the Brownian profile corresponding to the
same interval. The latter was combined with the diffusion constant $2\times
10^{-9}{\rm m}^2/ {\rm s}$ (see Fig.~\ref{fig4}). It is obvious that the
subdiffusive profile lags far behind the wider spread of the Brownian
counterpart, i.e., within the same time span, Brownian diffusion is more
efficient. Quantitatively, this corresponds to the ratio
$D_{2/3}/\left(D\Gamma(5/3)t^{1/3}\right)\approx 7.4t^{-1/3}{\rm s}^{1/3}$.

The subdiffusive diffusion profile sequence given in Fig.~\ref{fig8} is
juxtaposed to the theoretical curves in Fig.~\ref{fig11} (a). It is obvious
that the general trend follows the one displayed in Fig. \ref{fig8},
particularly that the spacing between successive curves becomes less pronounced
for increasing diffusion intervals. However, it can be realized that the
profile falls off too fast in comparison to the experimental result. We believe
that this is related to the fact that the investigated cluster is not an ideal
fractal since it is finite. This leads to corrections in the fractional model
that we heuristically introduce through a fudge factor as follows. Due to the
finiteness of the systems, there exist some channels from the left to the right
of the sample that enable a much faster, essentially Brownian, exchange with
the reservoir.

We therefore propose an ad hoc correction, namely that we have two additive contributions,
the anomalous profile plus a correction that is Brownian. That is, the
resulting profile becomes
\begin{equation}
\label{ff} C(x,\tau)=C_{\alpha}(x,\tau_{\alpha})+C_B(x,\tau_B),
\end{equation}
where $C_{\alpha}$ corresponds to the fractional solution with
$\tau_{\alpha}=(D_{2/3})^{3/2}t$ (see Tab.~\ref{tab1}) and $C_0=300$, and $C_B$
is given in terms of the Brownian solution given in Eq.~(\ref{csol}) with
$\tau=Dt$ and the fudge factor amplitude $C_0=10$. The latter was obtained from
the latest curve in Fig.~\ref{fig8} requiring that the value should be
approximately 9 at $x=40$mm. This is, of course, a rough and arbitrary choice
regarding the strong noise in the plot, and its purpose is only to demonstrate
the difference in the profile due to this procedure.

It is obvious from Fig.~\ref{fig11} (b) that even for the relatively small
ratio $C_B(0,\tau_B):C_{\alpha} (0,\tau_{\alpha})=1/30$, the profile is shifted
to higher values for larger $x$, and that the expression given in
Eq.~(\ref{ff}) seems to be a good approximation to the experimental result. It
might be argued that this measure violates the requirement that the mean
squared displacement should scale proportional to $t^{2/3}$. We plotted the
mean squared displacement corresponding to the modified profile,
Eq.~(\ref{ff}). It is obvious that the slope is approximately $2/3$, due to the
small relative contribution of the Brownian solution (note that the integral
determining the second moment of the Brownian contribution was cut off at
$x=80$mm since it has a non-negligible contribution for larger $x$). We expect
that the relative amplitude of the Brownian contribution decreases with
increasing system size, i.e., the finite size effects causing the necessity of
the Brownian correction should become smaller.

\section{Numerical evaluation based on the ordinary diffusion equation}

In addition to the analytical propagator treatment using the fractional
diffusion equation which describes the anomalous diffusion as an effective
result of the complex boundary conditions, we applied numerical finite volume
methods (FVM) to solve the ordinary diffusion equation (\ref{de}) for the
explicit  boundary conditions imposed by  the geometrical structure of the
finite model percolation cluster. The commercial software package FLUENT 5.5
(TM) provides the numerical basis for this sort of analysis.

The proton spin density distributions displayed in Fig.~\ref{fig6} were
considered as a solution of the ordinary diffusion equation for the pore space
of the percolation object and the initial condition of the experiment. Each
lattice site is covered by a grid of $5\times 5$ numerical unit cells. The
diffusion process occurring with the ordinary diffusion constant of water at
room temperature, $D=1.8\times 10^{-9}\, m^2/s$, was treated in a series of
$20\, s$ intervals. The proton concentration at the interface to the reservoir
was considered to be constant, $C(0,t)=1$, for all times.

Figure~\ref{fig14} shows the normalized second moment $ \langle x^2(t)\rangle$
evaluated from the spatial proton density distribution as a function of time
according to Eq.~(\ref{nmsd}). As a consequence of the tortuous diffusion pathways
in the percolation cluster, the time dependence turns out to be anomalous as
expected. That is
\begin{equation}
 \langle x^2(t)\rangle \propto t^{0.8}.
\end{equation}
The exponent value is coincides with the experimental result within the experimental
accuracy, but is slightly higher than the theoretical expectation for random-site percolation clusters
according to the Alexander/Orbach conjecture, Eq.~(\ref{aoc}).

The explanation is that there is  a finite contribution of undisturbed Brownian
diffusion trajectories through the pore space. The channel width relative to
the extension of the model system considered is not negligible as anticipated
in the theoretical percolation cluster model. Therefore, a finite fraction of
diffusion trajectories unaffected by the pore space restrictions contribute,
and the exponent will be correspondingly larger. Of course, this ``normal''
contribution is expected to vanish in the limit of infinite system sizes.

\section{Conclusions and discussion}

In this study, we have experimentally determined fractal  parameters of
random-site percolation clusters based on the structural properties of the
matrix, and grounded on the dynamic behavior of an isotopically labeled fluid
filled into the pore space.

Remarkably, anomalous diffusion was found for displacements far beyond the
correlation length, which is around 20 lattice constants at the occupation
probabilities considered in this study (see Fig.~\ref{fig3}). That is, the
normal diffusion limit in Eq.~(\ref{msd}) applies only extremely far above
$t_\xi$. This finding is in accordance with recent Monte Carlo simulations
\cite{poole96}, where the same conclusion was drawn.

Apart from the mean squared displacement considered  here, it is of interest to
compare the whole concentration profile with theoretical predictions. The
concentration profile in principle contains all information of the (anomalous)
propagator effectively determining the diffusion properties. This is shown with
the analytical propagator treatment based on the fractional diffusion equation
in Section~\ref{prop}. The experimental concentration profiles for water
diffusion in the model percolation clusters can be described very well on this
basis. In this way, the fractional diffusion equation formalism was verified
experimentally for the first time.

Additionally, we have evaluated the ordinary diffusion equation for the
boundary conditions given by the percolation model cluster. The experimental
data and the propagator results obtained from the fractional diffusion equation
coincide completely as fas as can be judged in the frame of the accuracy of the
diverse methods.

\acknowledgments This work was supported by the Deutsche
Forschungsgemeinschaft. We thank Hans Wiringer for
assistance in the course of this work, and Yossi Klafter
for inspiring discussions. Markus Weber kindly contributed the FVM program for the numerical solution of the ordinary diffusion equation. RM acknowledges financial support from the DFG within the Emmy
Noether programme.

\clearpage

\begin{figure}
\caption{Photograph of a section of a quasi  two-dimensional random site
percolation model object (topview) (a) and of an entire model object (cross
section) (b). The model object consists of several identical quasi
two-dimensional percolation clusters stacked on each other in order to improve
the signal intensity. The mechanical resolution of the fabrication process was
400~$\mu$m. The adjusted milling depth was constant between 1 to 2~mm in the
various objects produced. } \label{fig2}
\end{figure}

\begin{figure}
\caption{Proton spin density maps recorded  in a two-compartment sample
initially filled with bulk H$_2$O and D$_2$O gels (see the inset picture in
Fig.~\ref{fig4}). The echo time was $T_E=28$~ms, the repetition time was
$T_R=2$~s. The field of view in $x$ direction was 15~cm. The digital resolution
was 290~$\mu$m. The times indicate the span after contacting the two gels. The
white lines represent the isotopic interdiffusion profiles in the form of the
projection of the proton spin density on the $x$ direction.} \label{fig3}
\end{figure}

\begin{figure}
\caption{Comparison of theoretical  and experimental interdiffusion profiles
between bulk H$_2$O and D$_2$O gels. The data correspond to the two-dimensional
spin density maps shown in Fig.~\ref{fig3}. The solid lines represent fits of
Eq.~(\ref{profile}) to the experimental data. The fitted diffusion coefficients
are given in the inset table.} \label{fig4}
\end{figure}

\begin{figure}
\caption{Square of the position of  the proton concentration profile at half
height in bulk versus diffusion time. The experimental set-up consisted of two
compartments initially filled with bulk H$_2$O and D$_2$O gels as shown in the
inset picture. The largely asymmetric sizes of the H$_2$O and D$_2$O reservoirs
ensure a practically constant proton concentration at the entrance of the
D$_2$O compartment. The cross section of the D$_2$O compartment was
1.5~cm$\times$2~cm. The concentration profiles were obtained as projections of
two-dimensional spin density maps on the $x$ axis. The spin echo time was
$T_E=32$~ms, the repetition time was $T_R=0.7$~s. The field of view in $x$
direction was 7~cm. A digital resolution of 270~$\mu$m was adjusted. The total
acquisition time for a spin density map was 1~h.} \label{fig5}
\end{figure}

\begin{figure}
\caption{Evolution of two-dimensional  proton spin density maps ($256\times
256$ pixels) in a random site percolation object. The experimental setup is
schematically shown in the inset of Fig.~\ref{fig7}. The times indicate the
spans after attaching the H$_2$O gel compartment to the multi-stack percolation
model object (matrix size $100\times100$; $p-p_c=0.029$; $d_f=1.87$, porosity
of the percolating cluster $\rho=0.5352$ ). The time resolution given by the
image acquisition time was 26'. The echo time was $T_E=23$~ms, the repetition
time was $T_R=0.7$~s. The digital resolution of the maps is $230~\mu$m. The
white lines overlaid to the spin density maps represent the mean proton
concentration profiles (projections of the diffusion front on the main
diffusion direction).} \label{fig6}
\end{figure}

\begin{figure}
\caption{Square of the position of the  proton concentration profile at half
height in the percolation cluster shown and described in Fig.~\ref{fig6} versus
diffusion time. The inset shows the experimental set-up, where the percolation
cluster (white) initially is filled with D$_2$O. The largely asymmetric sizes
of the H$_2$O and D$_2$O reservoirs ensure a practically constant proton
concentration at the entrance of the D$_2$O compartment,  i.e. the percolation
cluster. The concentration profiles were obtained as projections of
two-dimensional spin density maps on the $x$ axis (see Fig.~\ref{fig6}). The
time resolution is 26'. The solid lines represent a fit of Eq.~(\ref{msd}) for
$t \ll t_\xi$. The fitted exponent parameter is $d_w=2.89$.} \label{fig7}
\end{figure}

\begin{figure}
\caption{Typical isotopic inter-diffusion  profiles directly measured with the
aid of one-dimensional imaging along the main diffusion direction. The
experimental set-up is schematically shown in the inset of Fig.~\ref{fig9}. The
spin-lattice relaxation time $T_1(x)$ was measured as a function of $x$ by
varying the repetition time $T_R$ from 0.2 to 12.2~s in 15 stps. The signal
intensity profiles were then corrected by multiplication with a factor $(1-\exp
\{ T_R/T_1(x)\})^{-1}$, so that any spatially dependent saturation effects were
eliminated. The spin-echo time was 20~ms, the field of view along the $x$
direction was 15~cm. A diffusion time resolution of 1~h, and a digital space
resolution of 290~$\mu$m were adjusted. } \label{fig8}
\end{figure}

\begin{figure}
\caption{Mean squared proton displacement as  a function of diffusion time in a
two-dimensional random site percolation object the template of which is shown
in the inset. The percolation cluster is initially ($t=0$) filled with D$_2$O.
The characteristic data of the percolation cluster are: matrix size $200\times
200$; occupation probability relative to the threshold value $p-p_c=0.030$;
fractal dimension $d_f=1.87$; porosity of the percolating cluster $\rho =
0.4845$.  The mean squared displacement data were evaluated from proton
concentration profiles recorded in the form of one-dimensional NMR images along
the $x$ axis. The proton concentration profiles were used to calculate the mean
squared proton displacement in the percolation cluster moiety. The solid line
represents a fit of Eq.~(\ref{msd}) for $t \ll t_\xi$. The fitted exponent
parameter is $d_w=2.86$. } \label{fig9}
\end{figure}

\begin{figure}
\caption{Comparison between the anomalous concentration profile $C_{2/3}
(x,\tau)$, left, and its Brownian analogue $C_1(x,\tau)$, for a diffusion time
interval $t= 180$~h. We used $D_{2/3}=1.3\times 10^{-8}{\rm m}^2/{\rm s}^{2/3}$
and $D=2\times 10^{-9}{\rm m}^2/{\rm s}$. The amplitude is $C_0=300$ in both
cases. The Brownian profile shows the much more efficient spread of the tracer
substance into the medium.} \label{fig10}
\end{figure}

\begin{figure}
\caption{(a) Anomalous concentration profile for $\alpha=2/3$, and with
$D_{2/3}=1.3\times 10^{-8}{\rm m}^2/{\rm s}^{2/3}$. The times increase away
from the origin, and are taken to be 20~h, 60~h, 100~h, 140~h, and 180~h,
corresponding to Fig.~\ref{fig8}. The inset shows a zoom into the plot range
$[0,25]$. The conversion between $t$ and the rescaled time $\tau$ is given in
Tab.~\ref{tab1}. (b) Corrected anomalous concentration profile,
Eq.~(\protect\ref{ff}), with the subdiffusive parameters $C_0=300$ and
$D_{2/3}=1.3\times 10^{-8} {\rm m}^2/{\rm s}^{2/3}$, and the Brownian diffusion
constant $D=2\times 10^{-9}{\rm m}^2/{\rm s}$. The fudge amplitude for the
Brownian contribution is 10.} \label{fig11}
\end{figure}

\begin{figure}
\caption{Mean squared displacement corresponding to the modified profile,
Eq.~(\protect\ref{ff}), $\log_{10}$-$\log_{10}$ scale (full line). The dashed
line represents the mean squared displacement corresponding to the fractional
result, Fig.~\ref{fig9}. Both lines are approximately of slope $2/3$.}
\label{fig13}
\end{figure}

\begin{figure}
\caption{Time dependence of the normalized mean squared displacement evaluated
from numerical solutions of the ordinary diffusion equation for the boundary
conditions given by the percolation model clusters (see Fig.~\ref{fig6}). The
numerical procedure is based on the finite volume method (FVM). The numerical
transient time resolution is $20\, s$. The time evolution is scanned in steps
of $\Delta t=3600\, s$. The data can be described by a power law $\langle
x^2(t) \rangle \propto t^{0.8}$.} \label{fig14}
\end{figure}

\clearpage

\begin{table}
\begin{tabular}{c|c|c}
$t$ [h] & $\left(D_{2/3}\right)^{3/2}t$ [$10^{-6}{\rm m}^3$] & $Dt$
[$10^{-4}{\rm m}^2$]\\
\hline\hline
20 & 0.11 & 1.44\\
60 & 0.33 & 4.32\\
100 & 0.56 & 7.20\\
140 & 0.78 & 10.10\\
180 & 1.00 & 12.96
\end{tabular}
\caption{Conversion between experimental time scale and rescaled ``time''
$\tau=Dt$ for Brownian and anomalous case. We use $D=2\times 10^{-9}{\rm m}
^2/{\rm s}$ and $D_{2/3}=1.3\times 10^{-8}{\rm m}^2/{\rm s}^{2/3}$.
\label{tab1}}
\end{table}


\begin{thebibliography}{99}
\bibitem{rem} We note that $K_{\alpha}$ is connected with the generalized
friction constant $\eta_{\alpha}$ through the generalized Einstein-Stokes
relation $K_{\alpha}=k_BT/(m\eta_{\alpha})$ \protect\cite{metzler00}. This
relation could be observed experimentally \protect\cite{amblard} and the use of
$K_{\alpha}$ therefore has a similar physical meaning as the traditional
diffusion constant $K\equiv D$ in Brownian motion.
\end{thebibliography}
\end{document}